\documentclass[amsmath,amssymb,floatfix,prl,twocolumn,pslatex]{revtex4}
\usepackage{graphicx}
\usepackage{dcolumn}
\usepackage{bm}

\begin{document}
\title{Concentration-dependent mobility in organic field-effect
transistors probed by infrared spectromicroscopy of the charge
density profile}

\author{A. D. Meyertholen, Z. Q. Li, D. N. Basov, and M. M. Fogler}


\affiliation{Department of Physics, University of California San
Diego, La Jolla, 9500 Gilman Drive, California 92093}

\author{M. C. Martin}

\affiliation{Advanced Light Source Division, Lawrence Berkeley
National Laboratory}

\author{G. M. Wang, A. S. Dhoot, D. Moses, and A. J. Heeger}

\affiliation{Institute for Polymers and Organic Solids and
Mitsubishi Chemical Center for Advanced Materials, University of
California, Santa Barbara}
\begin{abstract}
We show that infrared imaging of the charge density profile in
organic field-effect transistors (FETs) can probe transport
characteristics which are difficult to access by conventional
contact-based measurements. Specifically, we carry out experiments
and modeling of infrared spectromicroscopy of
poly(3-hexylthiophene) (P3HT) FETs in which charge injection is
affected by a relatively low resistance of the gate insulators. We
conclude that the mobility of P3HT has a power-law density
dependence, which is consistent with the activated transport in
disorder-induced tails of the density of states.
\end{abstract}



\date{\today}

\maketitle

Transport characteristics of organic semiconductors are important
for optimizing the performance of organic FETs. Due to substantial
disorder, polaronic, and interaction effects, transport mechanisms
in organics are complicated and standard model assumptions used
for inorganic devices may be unrealistic. For example, the
canonical model of the FET~\cite{Sze_book} assumes that the
mobility $\mu$ is independent of the areal carrier density $N$ in
the accumulation (or inversion) layer, while theoretical and
experimental investigations of organic conductors point to the
contrary~\cite{Brown_97, Vissenberg_98, Michels_05, Burgi_02}.
Unfortunately, the extraction of mobility from traditional
transport measurements is complicated by the nonlinearity of the
contact resistance, which may obscure the intrinsic mobility of
the organic semiconductor~\cite{Hamadani_04, Natali_07}.

Recently, we have reported a direct imaging of the charge
distribution in the accumulation layer of organic FETs using
infrared (IR) spectromicroscopy~\cite{Li_06}. In this Letter we
show that this contactless technique is complementary to
traditional DC transport measurements. Specifically, the carrier
density profile in the conducting channel obtained from the IR
scans can be used to extract the intrinsic behavior of the
density-dependent mobility, avoiding artifacts due to the
contacts.

As a case study, we investigated P3HT-based FETs in which an
inhomogeneous charge density profile was formed due to a
non-negligible leakage current through the gate dielectrics. Below
we show that the idealized model~\cite{Sze_book} fails to
correctly account for the charge density distribution observed in
these FETs. Instead, we find a good agreement with the model of a
density-dependent mobility,
\begin{equation}
\mu = {\rm const}\,\times N^\beta,\quad \beta = (E_* / k_B T) - 1
\geq 0, \label{eq:mu}
\end{equation}
which is derived theoretically below assuming transport is activated
and the density of states has an exponential tail with the
characteristic energy width $E_\ast$. Since $E_*$ depends on
disorder, exponent $\beta$ is not known precisely. From the fits to
our IR spectromicroscopy data we find $1 < \beta < 4$ at temperature
$T = 300\,{\rm K}$, where all the measurements are done. This
implies $E_* \sim 0.1\,{\rm eV}$, which agrees with other
experiments on P3HT~\cite{Tal_05}. The power-law
behavior~(\ref{eq:mu}) has been previously observed both in
P3HT~\cite{Burgi_02} and in other organic
conductors~\cite{Brown_97}. Thus, it is common to this class of
materials.

\begin{figure}
\begin{center}
\centerline{
\includegraphics[width=0.90\linewidth]{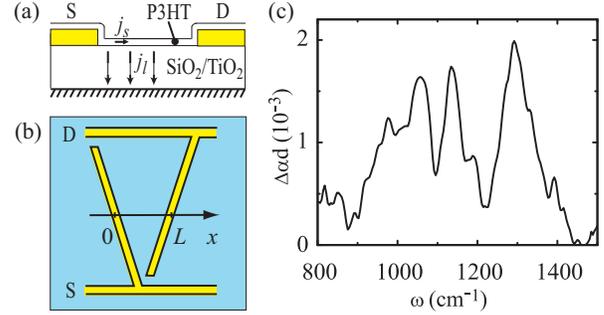}
}
\caption{(a) FET schematics (b) A sketch of the V-shaped
electrodes between which the IR scans were taken. (c) The
voltage-induced absorption spectrum $\delta \alpha d$ of one of
the FET devices studied. The sharp absorption peaks are the
IR-active vibrational modes of P3HT. \label{fig:device} }

\end{center}
\end{figure}

Let us proceed to the description of our experiment. The structure
of the FETs is shown in Fig.~\ref{fig:device}(a). In these
bottom-contacted devices, source and drain {Au} electrodes were
deposited on {SiO}$_2$/{TiO}$_2$/n-{Si} substrates followed by
the spin coating of a $4$--$6\,{\rm nm}$ thick P3HT film, which
served as the electronically active material. To examine length
scales associated with charge injection, we fabricated devices
with a V-shaped electrode pattern [Fig.~\ref{fig:device}(b)] where
the electrode separation varied from $0.01\,{\rm mm}$ at one side
to $4\,{\rm mm}$ at the other. To image the injected charges we
have employed the IR spectromicroscopy, which involved scanning a
focused IR beam $50$--$100\,\mu{\rm m}$ in diameter across the
region between the V-shaped electrodes and monitoring the changes
in the P3HT absorption spectrum at a given gate
voltage~\cite{Li_06}.

\begin{figure}
\begin{center}
\centerline{
\includegraphics[width=0.75\linewidth]{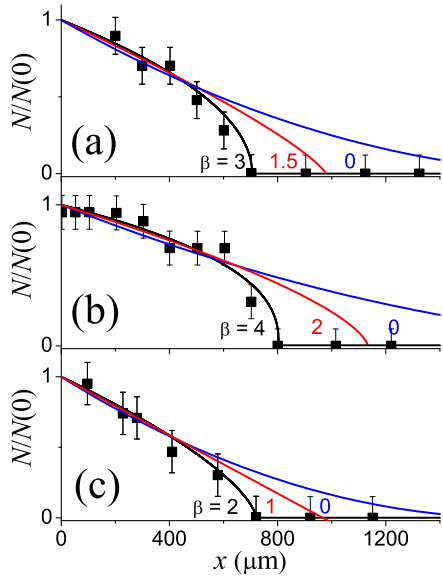}
}

\caption{Charge density profiles measured by the IR
spectromicroscopy in three nominally similar devices. The solid
lines are theoretical fits (see text). \label{fig:beta}}
\end{center}
\end{figure}

The spectroscopic ``fingerprints'' of the injected charges in P3HT are
the IR-active vibrational modes [Fig.~\ref{fig:device}(c)] and the
polaron band (a broad absorption maximum at $\omega \sim 3500\, {\rm
cm}^{-1}$, cf.~Fig.~2 of Ref.~\cite{Li_06}). From the oscillator
strength of the former, we computed the gate-induced charge density
$N(x)$ as a function of the distance $x$ from the source electrode.
(These results agreed within the experimental accuracy with $N(x)$
computed from the polaron band spectra, see Ref.~\cite{Li_06} for
details.) In Fig.~\ref{fig:beta}, we plot $N(x)$ normalized to its value
at $x = 0$ for three representative devices. Our main result is that the
injected charge density decays with the distance from the source and
vanishes at $x \sim 800\,\mu{\rm m}$. Below we show that the spatial
variation of the charge density contains valuable information on the
$N$-dependence of the mobility in accumulation layers of organic FET.

Indeed, in the experiment a non-negligible leakage current $j_l$
through the dielectric~\cite{Wang_04} was detected. The
corresponding effective resistivity of the dielectric layer
${\rho}_l = V / j_l$ was a function of $V$, the voltage difference
between the polymer and the gate. However, $\rho_l \approx {\rm
const}$ was observed~\cite{Wang_04} at $V
> 5\, {\rm V}$ and, for simplicity, we assume this to be the case. In the
gradual-channel approximation suitable for our FETs we obtain the
balance of surface (${\bf j}_s$) and leakage ($j_l$) currents as
follows:
\begin{equation}
-{\nabla} {\bf j}_s = {\bf \nabla} \left({\sigma} {\bf \nabla}
V\right)
                    = V / {\rho_l}.
 \label{eq:current}
\end{equation}
For the sheet conductivity $\sigma$ we adopt the form $\sigma =
\sigma_a + \sigma_r$, where $\sigma_a$ and $\sigma_r$ are the
sheet conductivities of the accumulation layer and the residual
mobile charges, respectively. Initially, let us make the
conventional assumptions that (i) $\mu \equiv \sigma_a / N = {\rm
const}$ (ii) $\sigma_r \ll \sigma_a$, and (iii) $V = e N / C$,
where $C = {\rm const}$ is the capacitance per unit area ($C
\approx 0.14\, \mu{\rm F}/{\rm cm}^{2}$ in our FETs). This entails
\begin{equation}
{\bf \nabla} \left({N}{\bf \nabla} N\right) \simeq {N} / (e\mu
\rho_l).
 \label{eq:N}
\end{equation}
Since the characteristic length of charge inhomogeneity is smaller
than the lateral extension of the source and drain electrodes in
the V-shaped region [Fig.~\ref{fig:device}(b)], it is permissible
to treat them as infinite metallic half-planes, $x < 0$ and $x >
L$, respectively. In this case, all variables depend only on $x
\in (0, L)$. Suppose first that $L$ is very large (the source and
drain are far apart), then the physically relevant solution of
Eq.~(\ref{eq:N}) is
\begin{equation}
 N(x) \simeq N(0) [1 - ({x}/{x_{\ast}})]^2,\quad
 x < x_\ast \equiv \sqrt{6 \sigma(0) \rho_l}.
\label{eq:sln}
\end{equation}
Interestingly, it predicts that the charge injection has a definite
cutoff length $x_\ast$. This occurs because of the quadratic
nonlinearity of Eq.~(\ref{eq:N}). Treating $x_\ast$ as an adjustable
parameter, we plot Eq.~(\ref{eq:sln}) in Fig.~\ref{fig:beta} (the curves
labeled $\beta = 0$). Alternatively, we can estimate $x_\ast$ based on
$\sigma$ and $\rho_l$ deduced from the DC transport
measurements~\cite{Wang_04, Li_06}, $\sigma \sim 10^{-7} \Omega^{-1}$
and $\rho_l \sim 10^5\, \Omega\, {\rm cm}^2$, which yields $x_\ast \sim
2500\,\mu{\rm m}$. Comparing with Fig.~\ref{fig:beta}, we see that the
observed $x_*$ is significantly smaller. Furthermore, instead of a more
gradual decrease in the conventional theory, in experimental data $N(x)$
exhibits an abrupt drop. This suggests that the degree of nonlinearity
of the actual current balance equation~(\ref{eq:current}) is higher than
quadratic.

The field-dependence of $\mu$ is probably not the main source of
this effect because we study P3HT of a rather high mobility $\mu =
0.05$--$0.12\, {\rm cm}^2 / {\rm V} {\rm s}$, where non-Ohmic
behavior is not as pronounced as in low-mobility
organics~\cite{Vissenberg_98, Michels_05, Hamadani_04, Burgi_02},
and also because $T$ was relatively
high~\cite{Comment_on_non-Ohmic}. On the other hand, there is a
natural reason for the $N$-dependence of $\mu$ [conveyed by
Eq.~(\ref{eq:mu})]. At densities we deal with, $N < 10^{13}\,{\rm
cm}^{-2}$, the system approaches the insulator-to-metal transition
region~\cite{Dhoot_06}. The chemical potential $\zeta$ presumably
resides in the density-of-states tail of disorder-induced traps
but at temperatures of interest such traps are shallow (a few $k_B
T$ deep). Therefore, transport is dominated by the activation of
carriers to either the mobility edge or the so-called transport
energy level~\cite{Vissenberg_98, Michels_05} and it is
characterized by a modestly large activation energy $|\zeta|$,
i.e., $\sigma_a \propto \exp({\zeta}/ k_B{T})$ (the mobility edge
is taken to be the energy reference
point)~\cite{Comment_on_traps}. Because of this exponential
relation between the conductivity and $\zeta$, the dependence of
$\zeta$ on $N$ is amplified in $\sigma_a$, producing the sought
nonlinear effect.

It is possible to express $\sigma_a$ directly as a function of $N$
if the density of states $g(E)$ is known. In conducting polymers
$g(E)$ is often modeled by Gaussian or exponential
forms~\cite{Vissenberg_98}, which are observed
experimentally~\cite{Tal_05}.  In either case we obtain $N(\zeta)
= \int dE g(E)f(E, \zeta) \simeq N_t \exp(\zeta / E_{\ast})$, up
to logarithmic factors.  Here $f(E, \zeta)$ is the Fermi-Dirac
distribution function, $E_{\ast}
> k_B T$ is the characteristic energy width of the band tail, and
$N_t$ is the total number of states per unit area in this tail. We
see that $\zeta \simeq E_\ast \ln (N / N_t)$, leading to $\sigma_a
\simeq \sigma_\ast (N / N_t)^{\beta + 1}$ with $\beta + 1 =
E_{\ast} / k_B T$. Consequently, the mobility is $\mu(N) \equiv
\sigma_a / e N \propto N^\beta$, in agreement with
Eq.~(\ref{eq:mu}). We note that a nearly identical $\mu(N)$
dependence also follows from the hopping transport
model~\cite{Vissenberg_98, Michels_05}. Equation~(\ref{eq:mu}) was
also postulated in Ref.~\cite{Natali_07}. Substituting $\sigma
\propto N^{\beta + 1}$ into Eq.~(\ref{eq:current}) but still
neglecting $\sigma_r$, we obtain
\begin{equation}
N \simeq N(0) \left(1 - \frac{x}{x_{\ast}}\right)^{\frac{2}{\beta
+ 1}}\!\!\!, \:\:\: x_{\ast} = \frac{\sqrt{2\left(\beta + 3\right)
\sigma(0) \rho_l}}
                {\beta + 1}
\label{eq:sln2}
\end{equation}
with Eq.~(\ref{eq:sln}) being recovered for $\beta = 0$. Three
representative curves computed according to Eq.~(\ref{eq:sln2})
are plotted in each panel of Fig.~\ref{fig:beta}. Clearly, for all
of the three FETs studied, one can find $\beta$s somewhere in the
range $1 < \beta < 4$ which give a significantly better fit to
experiment than $\beta = 0$, both in the functional form and in
the value of $x_*$. Apparently, Eq.~(\ref{eq:mu}) is more suitable
for modeling the charge injection profile in a ``leaky" FET.

For future reference we mention that a refinement of the model
including the residual conductivity,
\begin{equation}
   \sigma = \sigma_\ast (N / N_t)^{\beta + 1} + \sigma_r.
\label{eq:V}
\end{equation}
predicts the injected charge density profile of the form
\begin{align}
\frac{x}{x_\ast} &= - \sqrt{u + a}
                 + \frac{2 \sqrt{a}}{\beta + 3}\,
                   \ln \frac{\sqrt{u + a} + \sqrt{a}}{\sqrt{u + a} - \sqrt{a}}
                 + {\rm const},
\label{eq:injection_with_r}\\
u &\equiv [N(x) / N(0)]^{\beta + 1},\quad
 a \equiv (\beta + 3) \sigma_r / [2 \sigma_a(0)].
\label{eq:u}
\end{align}
Accordingly, $N(x) \propto \exp(-x / \sqrt{\sigma_r \rho_l})$ at
$x > x_*$, i.e., it has the exponential decay instead of the
abrupt threshold predicted by~Eq.~(\ref{eq:sln2}). However, since
$\sigma_r / \sigma(0) \sim 10^{-4}$--$10^{-3}$ (the on-off
ratio)~\cite{Wang_04}, the pre-exponential factor of this tail is
very small, producing apparently no signature in our experiment,
see Fig.~\ref{fig:beta}.

Finally, let us also discuss the $L < x_\ast$ case where the
injected charge is able to reach the drain (as in {SiO}$_2$
devices we studied previously~\cite{Li_06}). Here $N(x)$ depends
on the bias $V_{SD} \equiv V(0) - V(L)$. Also, the right-hand side
of Eq.~(\ref{eq:current}) can be set to zero because we can
normally neglect the leakage current $j_l$ compared to the
source-drain current $j_s$. Equation~(\ref{eq:current}) remains
solvable~\cite{Comment_on_saturation} and yields the algebraic
equation for $N(x)$,
\begin{equation}
\frac{\sigma_\ast N_t}{\beta + 2}
\left(\frac{N}{N_t}\right)^{\beta + 2} + \sigma_r N = \frac{C
j_s}{e} (x_p - x), \label{eq:Vpinch1}
\end{equation}
where $x_p$ is the integration constant. At $V(L) \rightarrow 0$,
we have $x_p \rightarrow L$ and, except very near the drain, the
first term on the left-hand side dominates. In this case the
charge profile is very similar to that of the injection case,
Eq.~(\ref{eq:sln2}), except $x_\ast$ is replaced by $L$ and the
power-law exponent $2 / (\beta + 2)$ is changed to a smaller
number $1 / (\beta + 2)$. Although we were not able to verify this
formula by our technique (in part, due to a finite spatial
resolution $\sim 100 \,\mu{\rm m}$) we note that B\"urgi {\it et
al.\/}~\cite{Burgi_02} have studied this regime in similar FET
devices using scanning potentiometry. These authors were able to
fit their data to the formula obtained by setting $\sigma_r = 0$
and $\beta = 1$ in our
Eq.~(\ref{eq:Vpinch1})~\cite{Comment_on_Burgi}.

In conclusion, when the charge profile in the organic
semiconductor is strongly inhomogeneous, as in the charge
injection~\cite{Li_06} or the current saturation~\cite{Burgi_02}
regimes, its modeling should include the density dependence of the
carrier mobility. Our experiment is consistent with the power
law~(\ref{eq:mu}), which has a natural physical motivation. Our
measurements and the method of analysis are free from problems
introduced by the contact resistance.


Work at UCSB was supported by the NSF grant DMR-0602280. The UCSD team
was supported by the NSF grant ECS-0438018 and UCSD ACS. We thank
D.~Natelson for comments on the manuscript.




\begin{thebibliography}{17}
\bibitem{Comment_on_non-Ohmic} No significant field-dependence of $\mu$
was found experimentally at such temperature~\cite{Burgi_02,
Wang_04}. The nonlinearity of the contact resistance can still be
significant~\cite{Burgi_02}; however, it has no effect on our
contactless measurements.
\bibitem{Comment_on_traps} By means of repeated capture and release from
shallow traps, charge carriers in our experiment travel
macroscopic distance (almost a ${\rm mm}$) away from injecting
electrodes and the decay of $N(x)$ is detected chiefly because we
intentionally work with a large-area FET with a non-negligible
leakage current.
\bibitem{Comment_on_saturation} Deep in the current saturation regime
(where $V(0)$ and $V(L)$ are of opposite sign) one cannot treat
the problem as one-dimensional, and $\sigma_r$ becomes position
dependent, see, e.g., Ref.~\cite{Brown_97}. We do not consider
this case here.
\end{thebibliography}
\end{document}